\documentclass[lettersize,journal]{IEEEtran}

\usepackage[utf8]{inputenc}

\usepackage{array}
\usepackage{multirow}

\usepackage{xurl}

\usepackage{graphicx}
\graphicspath{ {figures/} }

\begin{document}

\title{A Survey of AI-Related Cyber Security Risks and Countermeasures in Mobility-as-a-Service}

\author{Kai-Fung Chu,~\IEEEmembership{Member,~IEEE},
Haiyue Yuan,~\IEEEmembership{Member,~IEEE},
Jinsheng Yuan,
Weisi Guo,~\IEEEmembership{Senior Member,~IEEE},
Nazmiye Balta-Ozkan,~\IEEEmembership{Member,~IEEE},
Shujun Li,~\IEEEmembership{Senior Member,~IEEE}%
\thanks{This work was supported by the Engineering and Physical Sciences Research Council (EPSRC), part of the UK Research and Innovation (UKRI), as part of the research projects ``MACRO -- Mobility as a service: MAnaging Cybersecurity Risks across Consumers, Organisations and Sectors'' (EP/V039164/1), and ``TAS-S: Trustworthy Autonomous Systems: Security'' (EP/V026763/1).}%
\thanks{Kai-Fung Chu is with the Department of Engineering, University of Cambridge, Cambridge CB2 1PZ, UK. (e-mail: kfc35@cam.ac.uk)}
\thanks{Haiyue Yuan and Shujun Li are with the Institute of Cyber Security for Society (iCSS) \& School of Computing, University of Kent, Canterbury CT2 7NP, UK.}
\thanks{Jinsheng Yuan and Weisi Guo are with the School of Aerospace, Transport and Manufacturing, Cranfield University, Milton Keynes, MK43 0AL, UK.}
\thanks{Nazmiye Balta-Ozkan is with the School of Water, Energy and Environment, Cranfield University, Milton Keynes, MK43 0AL, UK.}%
}

\markboth{IEEE Intelligent Transportation Systems Magazine, Vol.~X, No.~Y, Xxx 2024}{Chu \MakeLowercase{\textit{(et al.)}}: AI-Related Risks and Countermeasures in MaaS}


\maketitle

\begin{abstract}
Mobility-as-a-Service (MaaS) integrates different transport modalities and can support more personalisation of travellers' journey planning based on their individual preferences, behaviours and wishes. To fully achieve the potential of MaaS, a range of AI (including machine learning and data mining) algorithms are needed to learn personal requirements and needs, to optimise journey planning of each traveller and all travellers as a whole, to help transport service operators and relevant governmental bodies to operate and plan their services, and to detect and prevent cyber attacks from various threat actors including dishonest and malicious travellers and transport operators. The increasing use of different AI and data processing algorithms in both centralised and distributed settings opens the MaaS ecosystem up to diverse cyber and privacy attacks at both the AI algorithm level and the connectivity surfaces. In this paper, we present the first comprehensive review on the coupling between AI-driven MaaS design and the diverse cyber security challenges related to cyber attacks and countermeasures. In particular, we focus on how current and emerging AI-facilitated privacy risks (profiling, inference, and third-party threats) and adversarial AI attacks (evasion, extraction, and gamification) may impact the MaaS ecosystem. These risks often combine novel attacks (e.g., inverse learning) with traditional attack vectors (e.g., man-in-the-middle attacks), exacerbating the risks for the wider participation actors and the emergence of new business models.
\end{abstract}

\begin{IEEEkeywords}
Mobility-as-a-Service, transport, machine learning, cyber security, privacy, business model, low carbon MaaS
\end{IEEEkeywords}

\section{Introduction}

Mobility-as-a-Service (MaaS) is an innovative mobility concept that aims to integrate various transport modes into a single platform~\cite{hietanen2014mobility}, providing passengers with real-time traffic information, journey planning, booking, and bundle offers across different transport operators. Besides addressing the fluctuating transport demand, MaaS has the potential to integrate novel intelligent transportation systems applications, such as predicting future traffic information~\cite{chu2019deep}, enabling ride-sharing~\cite{agatz2012optimization}, and increasing vehicle utilisation~\cite{chu2021joint}. These highlight how MaaS, with today's existing fossil-fuel dominated infrastructure systems, has the potential to address various transportation challenges, including traffic congestion, air pollution, and energy consumption. As more and more electric or fuel cell buses, vehicles and trains are integrated to decarbonise transport systems~\cite{Iea}, the future `\textit{low carbon MaaS}' can provide additional benefits beyond transport systems for the operation and planning of energy systems by controlling when, how much and where to charge them. These benefits can help a number of actors in energy systems and balancing and ancillary services markets, including suppliers, aggregators, distribution and transmission system operators. Coupled with paradigm shift from a vehicle ownership-based system towards an access-based one, whether low carbon or not, these opportunities highlight the level of intelligence that future transport systems may aspire to have.

MaaS was first formally defined by Hietanen in 2014~\cite{hietanen2014mobility}, who described MaaS as a mobility distribution model fulfilling users' transportation demands via a single interface of multiple service providers. It combines different transport services to a tailored mobility package, similar to a monthly mobile phone contract. Since then, more and more detailed features have been supplemented to this concept. Consumers may settle the payment by a seamless ``pay-as-you-go'' choice~\cite{goodall2017rise}, or personalised bundles option~\cite{atasoy2015concept}. To enable those features, MaaS relies on Information and Communication Technologies (ICTs) to integrate the information between consumers and providers~\cite{nemthanu2016mobility} and the Internet of Things (IoT) to handle the connection between physical components~\cite{sherly2015internet}. With the information access to various components in MaaS, an intelligent coordinator can leverage the available information to optimise the travel plan and provide consumers with the most efficient, convenient, and cost-effective transportation modes. Heavy reliance of MaaS on big data means that ICT including artificial intelligence (AI) technologies can enable its credibility and security by managing privacy and security threats. Those threats can be approximately divided into two types depending on where the attack surface lies (data and algorithmic levels).

For the data level, personal data privacy has become a big concern with the advancement of MaaS~\cite{huang2022listening}. Since most MaaS systems use a centralised mobility system to collect and process vast amounts of personal information, they are vulnerable to various data attacks, such as identity theft, unauthorised access, and data manipulation. That sensitive personal information of MaaS consumers may leak to threat actors such as developers, service administrators, and managers~\cite{callegati2018cloud}. For the algorithm level, the algorithms and systems used by MaaS can be vulnerable to attacks, such as adversarial attacks, data poisoning, and model extraction. Adversarial attacks can manipulate input data to cause incorrect or inappropriate outputs. Data poisoning attacks~\cite{steinhardt2017certified} can insert malicious training data to alter the model's behavior. Model extraction can steal the intellectual property of an AI model by collecting data through query access to the model.

\begin{figure}[!htb]
\centering
\includegraphics[width=\linewidth]{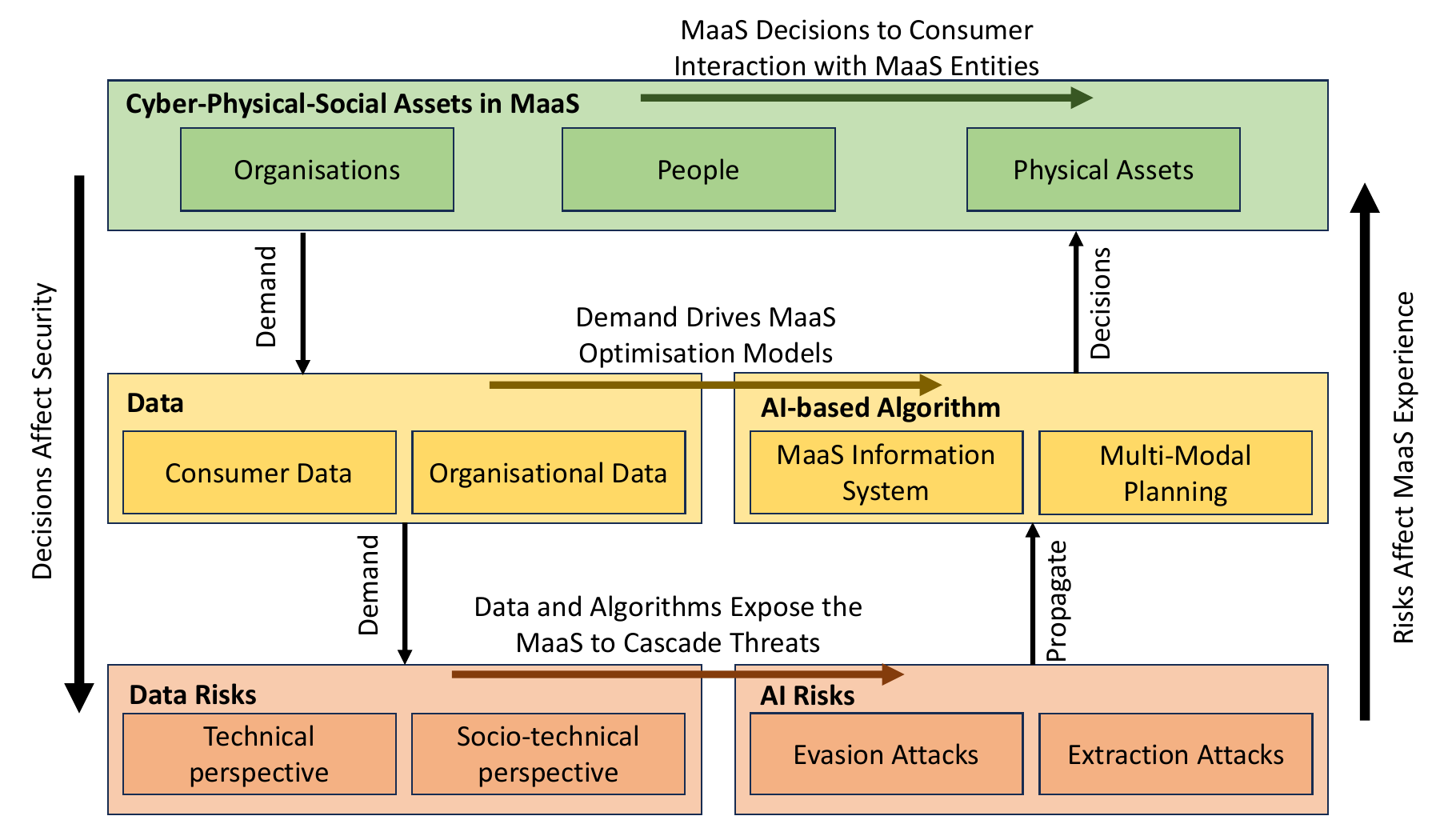}
\caption{In the MaaS ecosystem, cyber security risks manifest through diverse threats originating from adversarial entities, both within and outside the MaaS framework. These risks extend beyond conventional cyber assets to encompass socio-physical assets, such as individuals, organisations, and the data and AI assets under their purview. The complexity of this system underscores the multifaceted nature of data and AI risk generation and propagation. Decisions made by both human actors and automated systems, spanning individuals and organisations, play pivotal roles in shaping the dynamics of cyber security risks. Moreover, these risks transcend organisational boundaries and sectoral domains, permeating throughout the interconnected landscape of the MaaS ecosystem.}
\label{fig:overview}
\end{figure}

To ensure a safe and trustworthy MaaS ecosystem, a mature and secure software system that uses various defence mechanisms, such as input validation, outlier detection, and model watermarking, is necessary for the intelligent scheduler to connect operators and passengers, manage traffic information, and optimise passengers' journey queries and system resources, to maintain the service quality. Figure~\ref{fig:overview} outlines the relationship of the MaaS assets and cyber security risks. However, there is no existing survey for reviewing the state-of-the-art research work done for the cyber security and MaaS. The author of~\cite{maas2022literature} conducted a systematic literature analysis on MaaS with a short discussion in data security and resilience. Another survey~\cite{Butler-L2021} covers the barriers and risks of MaaS, which contributes a shallow summary of the cyber security risk out of other barriers such as collaboration, business support, coverage, and shared vision. On the other hand, detailed cyber security surveys were conducted for other systems. For example, a detailed survey on cyber security of autonomous mobility systems is presented in~\cite{zou2021cyber}, which focuses on the system related to autonomous vehicles. Researchers are interested in the cyber security of connected vehicles~\cite{olovsson2022future}. Hence, a cyber security survey for both connected and autonomous vehicles is presented in~\cite{sun2021survey}. Another cyber security survey was conducted for railway cyber-physical system~\cite{wang2022cyber}. Given the promising benefits of MaaS, the cyber security risks should be clearly identified such that the corresponding countermeasures can be implemented in the MaaS ecosystem. In this paper, we first review how AI and computational solutions can be integrated into the MaaS ecosystem. Then, we review the current cyber security risks and countermeasures in both data and AI aspects. Finally, we review the impact of AI-related risks on MaaS business models and discuss the future trends of MaaS.

In Section~\ref{system_planner_section}, we first explore the latest trends and technologies in the MaaS system planner, which is one of the data, algorithm, and risks hot spots, to understand the cyber security aspects of the MaaS ecosystem. In Section~\ref{data_section}, we review the data attack vectors and defence mechanisms related to the MaaS system. In Section~\ref{section:review_AI}, we highlight the risks in the state-of-the-art AI technologies and common practices for those AI risks. In Section~\ref{business_section}, we discuss the impact of the cyber security risks and countermeasures on the MaaS business ecosystem. Finally, we conclude the paper in Section~\ref{conclusion}.

\section{Design of MaaS System Planner}
\label{system_planner_section}

\begin{figure}[!htb]
\centering
\includegraphics[width=\linewidth]{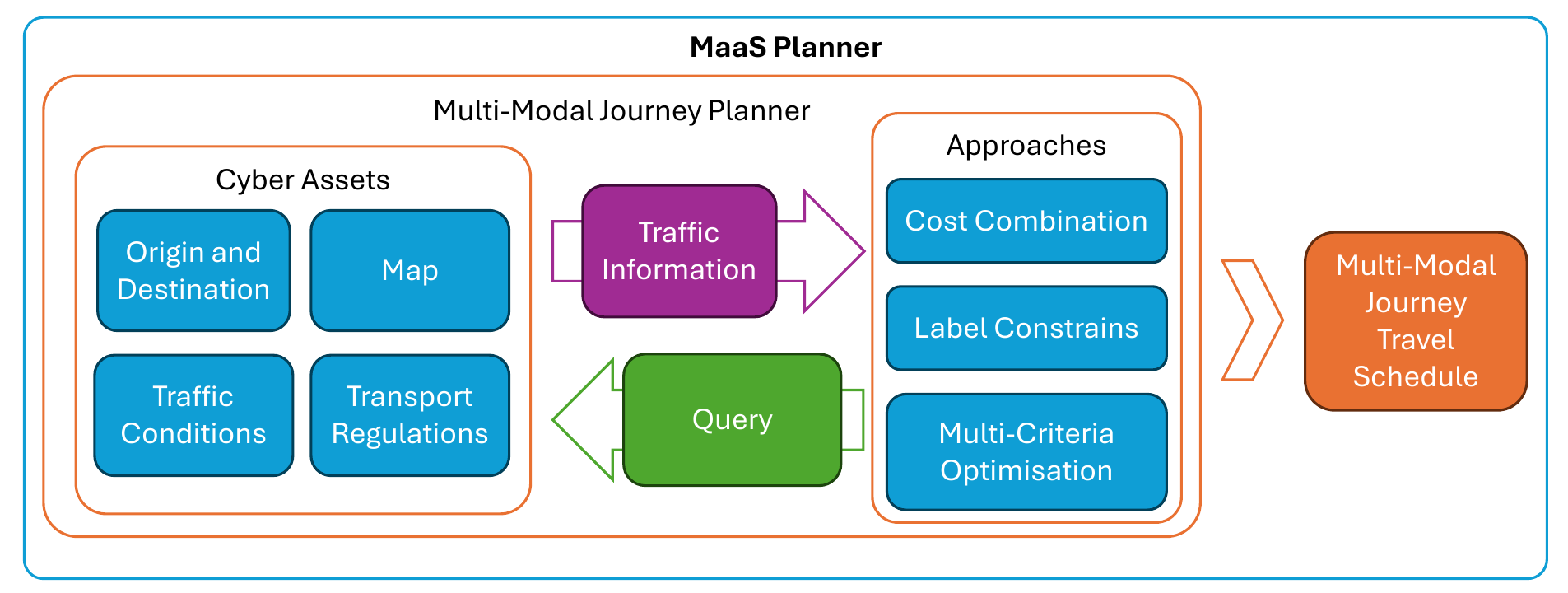}
\caption{A MaaS planner determines multi-modal journey using cost-combining, label-constrained, and multi-objective optimisation approaches based on the cyber assets of multiple service providers.}
\label{fig:maas_planning}
\end{figure}

The MaaS system planner is a crucial component in the daily operations of the service, as it is responsible for decision-making processes such as suggesting journeys, arranging transport schedules, and determining bundle pricing. Therefore, the quality of the method the planner employs heavily influences the effectiveness of the transportation system and consumer utility. Designing such a planner can be a challenging task, and a loosely organised solution could lead to the diminished utility for both transport operators and consumers. As a result, significant efforts have been put into designing the MaaS system planner. Figure~\ref{fig:maas_planning} outlines the framework of a MaaS planner, particularly for the multi-modal journey planning problem. The rest of this section presents the review of state-of-the-art research work of the MaaS system planner, including journey planning and the multi-model one. We also identify the future trends in this research direction at the end of this section.

\subsection{Journey Planning}

One of the most critical functions of a MaaS planner is journey planning, which provides consumers with information and guidance about the efficient and convenient travel experience. Finding the most efficient route from origin to destination in a transportation network has a long history, and it is formalised as a path planning problem~\cite{few1955shortest, dreyfus1969appraisal, chu2019dynamic}. It can be described as the problem of determining an optimal path over a graph, which comprises vertices that are connected by edges. Typically, the optimal path refers to the path with the minimal cost between the origin and destination. Various algorithms, such as Dijkstra's shortest path algorithm~\cite{dijkstra1959note}, A* search algorithm~\cite{hart1968formal}, and other variants~\cite{hilger2009fast, geisberger2012exact}, have been developed to solve the problem efficiently. To apply to transportation services with timetables, these static graph-based structure algorithms have been extended to handle the temporal nature of the problem. There are three types of time-related models: time-expanded, time-dependent, and frequency-based models, which are described as follows.
\begin{itemize}
\item Time-expanded models~\cite{schulz2000dijkstra, pyrga2008efficient} denote public transport stops and time events as separated stop nodes and time nodes. The time nodes are connected to the corresponding stop nodes. For example, a time node can define whether the departure of a vehicle is before the arrival of the other. Although the number of nodes and edges may scale up in these models, journey planning and different fast variant algorithms can be applied to the models.

\item In time-dependent models~\cite{brodal2004time, bast2016route}, public transport stops are represented by nodes. An edge between two nodes exists if the transport service operates between the stops. The weights of these edges usually represent the time cost between the stops. Compared to the time-expanded model, the time-dependent model could have a smaller graph size. However, the dynamic weights could be challenging for journey-planning algorithms. For example, a dynamic shortest path problem has been designed in~\cite{sever2018dynamic} to find a path with a minimum travel time from origin to destination using both historical and real-time information.

\item Frequency-based models~\cite{bast2014flow} store departure times, intervals, and frequencies rather than time-dependent functions. It uses the regularity of public transportation networks to compress the graph and decrease query times.
\end{itemize}
In terms of algorithms for those models, the time-expanded models can be pre-processed better, while time-dependent and frequency-based models rely on advanced algorithms to deal with the complex models. Moreover, the uncertainty in the transportation network is another factor that requires a dynamic updatable graph based on the real-time traffic state. Recently, more and more new approaches that employ real-time information have been integrated into the planner algorithm. For example, a journey planner can incorporate the prediction of future traffic conditions~\cite{liebig2017dynamic}.

\subsection{Multi-Modal Journey Planning}

In MaaS, the journey could be composed of multiple modes of transport to reach the destination over the multi-modal transport network, including fixed modes (public transport), dynamic modes (car-sharing and ride-sharing), and even automated in the future~\cite{he2020concept}. The journey planning's complexity is compounded by the various constraints and limitations of different services, such as geographical boundaries, temporal limitation, and availability of transport modes. Hence, MaaS requires a more capable multi-modal journey planning algorithm to address the challenges.

The naive approach to obtaining a multi-modal graph is by first building an individual graph for each transport mode, and then merging the involved graphs into a single multi-modal one~\cite{delling2009accelerating, yu2012advanced}. The multi-modal graph can combine different graphs, such as time-independent and time-dependent graphs. After the merger, the problem can be solved by a planning algorithm. In addition to this naive approach, we will discuss three different approaches to the multi-modal journey planning problem: cost-combining, label-constrained, and multi-objective optimisation~\cite{bast2016route}.

\subsubsection{Cost-Combining Approaches}

Since the problem involves the combination of different transport modes, one can add a penalty to the objective function to represent the introduced transition cost of transferring from one mode to another. In~\cite{modesti1998utility}, Modesti and Sciomachen presented an approach based on the classical shortest path problem on a network representing the urban multi-modal transport system, including unrestricted walking, unrestricted car travel, and public transit. The journeys are obtained by optimising a linear combination of criteria, such as cost, travel time, and user preferences on the transport modes. Aifadopoulou et al.~\cite{aifadopoulou2007multiobjective} proposed to solve the optimal journey problem using linear programming over multi-modal transportation networks. Antsfeld et al.~\cite{antsfeld2012finding} suggested using a linear utility function that incorporates travel time, ticket cost, and inconvenience of transfers. To merge the fixed and dynamic modes with the fuzzy and flexible nature, Huang et al.~\cite{huang2019multimodal} considered the concept of drive-time areas and points of action, which can merge them better while maintaining the flexibility of the dynamic modes. Pantelidis et al.~\cite{pantelidis2020many} formulated the traffic assignment in MaaS as stable matching of multiple transport operators and passengers. This matching aims to allocate costs and determine prices based on the route choices of passengers and the service choices of operators. Xu et al.~\cite{xu2021online} modelled the concept of congestive capacity on the route choice model, where link capacities depend on traffic flows rather than link costs. They proposed a method to obtain unique shadow prices for congestive capacity in a multi-modal transport network to capture the structural effects of flows on capacities and the resulting impacts on route choice utilities. The aim is to verify the capability to capture congestion effects on capacities. Ma et al.~\cite{ma2023dynamic} investigated dynamic bus-routing, integrating on-demand services with real-time passenger demand, utilising a two-stage stochastic programming model to optimise vehicle travel time cost and minimise the penalty for rejected requests.

\subsubsection{Label-Constrained Approaches}

In label-constrained approaches, the weights of the graph are labelled by an alphabet $\Sigma$ that denotes the modes of transport, and a language $L$ is specified in the shortest path problem such that journeys have to obey pre-defined constraints related to the modes of transport~\cite{barrett2000formal}. If the languages in the problem are regular, there are techniques to obtain tractable solutions~\cite{barrett2002classical, barrett2008engineering}. The label-constrained shortest path problem can be sped up by using an approach proposed by Delling et al.~\cite{delling2009accelerating}, which can handle hierarchical languages that allow constraints such as restricting walking and car travel from the beginning to the end of the journey. Khani~\cite{khani2019online} developed an online shortest path algorithm and a label-correcting solution algorithm in schedule-based transit networks with stochastic vehicle arrival times.

\subsubsection{Multi-Objective Optimisation Based Approaches}

Although label constraints are valuable in determining possible journeys, there are drawbacks in computing the shortest path with label constraints. Firstly, the characteristics of each transport network must be known in order to set the constraints. Secondly, the computed journeys are limited to certain transport mode combinations, while combining the modes differently is not computed. Therefore, multi-objective optimisation can be considered to determine diverse solutions.

In~\cite{martins1984multicriteria}, two algorithms are presented for solving the multi-objective shortest path problem, proving that any pair of non-dominated paths can be connected by non-dominated paths. Zografos et al.~\cite{zografos2008algorithms} formulated the journey planning problem on a multi-modal time-schedule network with time windows and time-dependent travel times that minimise a set of criteria, including the total travel, walking, and waiting time and the number of interchanges. Bast et al.~\cite{bast2013result} proposed a method called Types aNd Thresholds for identifying significant journeys in the non-dominated solution. It is based on a set of simple axioms that summarise what a majority of consumers as unreasonable journeys. The planner developed by Atasoy et al.~\cite{atasoy2015concept} provides an optimised menu of travel options complying with seat capacity and committed time schedule constraints to passengers. The model considers the trade-off between consumer surplus and operator profit to improve passenger satisfaction. Song et al.~\cite{song2021whole} formulated the whole-day multi-modal journey planning problem that considers user-specific modal preferences. The problem aims to minimise the total travel and waiting times and the transfer number in a day. Chu et al.~\cite{chu2023deep} introduced a novel approach for multimodal journey planning in MaaS that considers diverse passenger preferences and dynamic behaviours by modelling passenger experience as a Markov process and formulating a multi-objective optimisation problem. Constraints such as the time window and park-and-ride demands are considered to produce customised journey planning. This method could also be extended to low-carbon MaaS systems by collating and analysing data in the location and capacity of charging stations, as well as temporal changes in local and national energy and ancillary services markets~\cite{ho2024mobility}.

\subsection{Future Trends and Challenges}

Modern algorithms are anticipated to continue leveraging inherent characteristics of road networks to enhance efficiency. Despite sporadic developments in geometry-based methodologies, established techniques are expected to maintain dominance owing to their superior efficacy. The integration of real-world data in experimental validation processes will remain crucial to ensure the fidelity of algorithmic models to actual production data and to challenge conventional assumptions. Moreover, algorithms validated through real-world applications are likely to be integrated into systems serving vast user bases. However, persistent challenges include the realisation of a global multimodal journey planner incorporating real-time data and personalised elements efficiently, suggesting ongoing advancements will be necessary to address these complexities. With the increasing reliance on data-driven and AI-based technologies in transport networks, ensuring their security and resilience against potential attacks becomes paramount. Therefore, it is essential to complement these algorithmic advancements with AI-related cyber security strategies, to safeguard the integrity and reliability of future transport infrastructures.

\section{Review of Data Attack Vectors and Defense}
\label{data_section}

\begin{figure}[!htb]
\centering
\includegraphics[width=\linewidth]{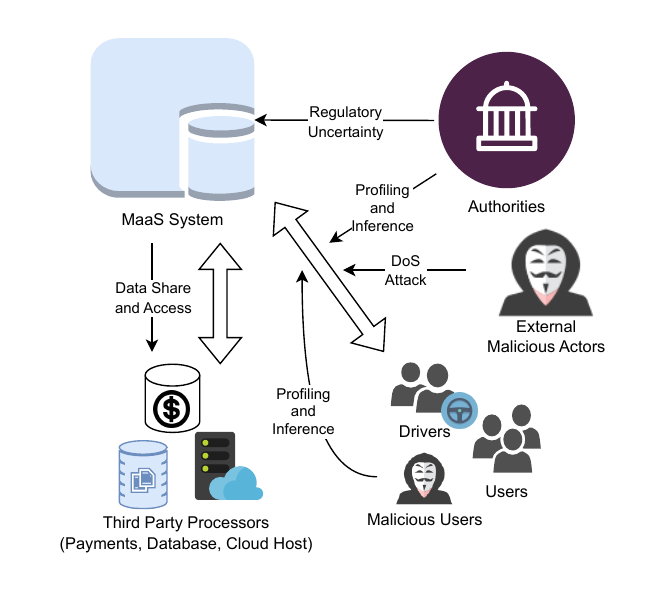}
\caption{Data attack vectors mainly consist of two categories, A. data privacy attacks and B. data-computer pipeline attacks. Data privacy risks include 1) profiling and inference and 2) third-party unauthorised access and data oversharing. Data-computer pipeline attacks include DoS (Denial of Service) attacks and attacks from socio-technical perspectives.}
\label{fig:data_attack_vectors}
\end{figure}

MaaS as an ecosystem integrates various transport services and possesses the capability and scalability to gather and analyse diverse data from customers and service providers, meaning that MaaS has the potential to accumulate a vast amount of personal and sensitive information, and involves a multitude of parties in the data processing chain. To this end, cyber security concerns and risks related to MaaS data are considerable, and many studies have been conducted to explore these from different perspectives (see Fig.~\ref{fig:data_attack_vectors}). We broadly define these in two categories, data privacy attacks and data-computer pipeline attacks, which we discuss in detail. The rest of this section presents our work of categorising privacy and security risks related to MaaS from different perspectives. We also identify several research gaps and future challenges, which will be presented at the end of this section to highlight the main takeaway messages and provide insights for future research directions.

\subsection{Data Privacy Attacks}

MaaS enables users to access comprehensive trip services—planning, booking, ticketing, payment, and real-time information within a single digital platform, eliminating the need for separate ticketing and payment operations. To achieve this, MaaS systems involve multiple parties that collect and process substantial amounts of personal and sensitive information. For instance, effective MaaS requires the combined time and location-specific travel behavioural data of individuals. Additionally, users must link their financial information for payment schedules, further increasing the amount of personal information. These facts collectively contribute to and raise potential user data and privacy concerns and increase privacy leakage risks, which can be broadly categorised into two groups based on related research work. The first one pertains to \emph{profiling and inference}, which deals with the collection of personal data, such as geo-location, mobile phone usage, and payment information, and how it can be used to profile and infer end users' behaviours and mobility patterns. The second group is \emph{third party unauthorised access and data oversharing}, which concerns the possible adversaries caused by third-party access via data breaches and/or unauthorised access to personal data collected by MaaS providers, and the unnecessary/unwanted oversharing of data among multiple parties. Both of these risks are amplified in \emph{low-carbon MaaS} systems as there are data exchanges between transport and energy systems to meet charging needs.

\subsubsection{Profiling and Inference}

\begin{figure}
\centering
\includegraphics[width=\linewidth]{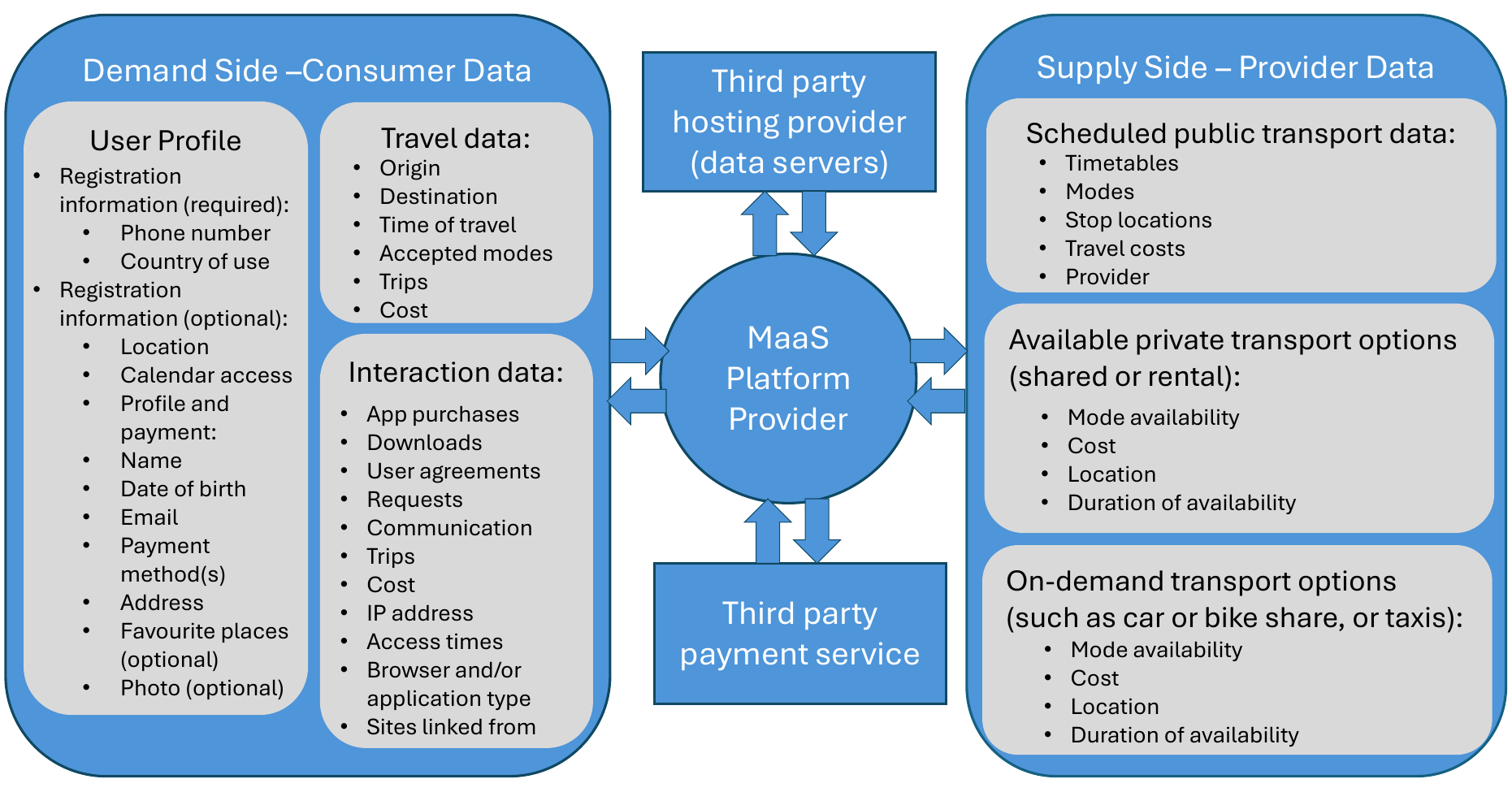}
\caption{High-level conceptualisation of a MaaS-style data ecosystem~\cite{Cottrill-C2020}}
\label{fig:data_scale}
\end{figure}

An example of the magnitude of personal data that can be collected by MaaS systems (see Fig.~\ref{fig:data_scale}) was given in a study~\cite{Cottrill-C2020}, where Cottrill reviewed the privacy policy of Whim, a MaaS app. The data that can be directly collected from consumers include basic personal details (e.g., name and telephone number), additional personal details (e.g., email address, home country and address, information on devices used, language, credit card details, etc.), and verification data (e.g., personal identity number, photo, or driving license details). Other data such as transaction information, location data, travel data can be also collected through the use of the app.

Barreto et al.~\cite{Barreto-L2018} conducted a study on utilising MaaS for urban mobility digitalisation and found that as a user shares personal data or registers with a MaaS service, the MaaS operator, such as the local authority managing a smart city, has the capability to infer the user's behaviours and mobility patterns, including their needs, interests, and possible transport means choices. Moreover, Costantini et al.~\cite{Costantini-F2019} stated that by analysing the mobility patterns extracted from MaaS users, it is possible to infer information related to certain health conditions, which can violate the user's privacy to some extent. Past research has also addressed the privacy risks of using collected geo-location data for profiling and inference. Costantini et al.~\cite{Costantini-F2019} pointed out that geo-location data contain a wealth of information that can create additional vulnerabilities, particularly when the data contain a specific user's sensitive destination information, such as a cult temple, the office of a syndicate or political party, a civil organisation, or a school for their child(ren). This may introduce legal concerns, as the collection and use of such personal data can raise legal compliance issues such as violation of one or more rights of data subjects. In addition, the combination of geo-location data with the corresponding time of use information can be highly valuable to companies operating in the retail and leisure sectors. However, the monetisation of such sensitive data raises serious privacy and ethical concerns, which should not be ignored~\cite{Cooper-P2019}.

While the privacy risks over MaaS consumers' data is relatively well established in the literature, data from MaaS service providers that can cause privacy risks and concerns is much less discussed. Drivers' schedule information~\cite{Belletti-F2017}, drivers' geo-location data~\cite{Callegati-F2018}, and drivers' performance records~\cite{Kong-Q2021} were all identified as containing sensitive information that can lead to re-identification risks and violations of location privacy. To mitigate potential risks, Belletti and Bayen~\cite{Belletti-F2017} proposed a constrained integer quadratic program-based framework that does not require personal availability constraints of drivers to be shared with their system. Kong et al.~\cite{Kong-Q2021} developed a blockchain-based solution to preserve the privacy of drivers that can be shared across MaaS operators.

\subsubsection{Third-Party Unauthorised Access \& Data Oversharing}


\begin{table}[!htb]
\centering
\resizebox{\linewidth}{!}{
\begin{tabular}{ccccccccc}
\cline{2-9}
\multirow{20}{*}{\rotatebox{90}{Switzerland}} & & & \multicolumn{6}{c}{Smart Mobility Policies} \\
\cline{4-9}
& \multirow{8}{*}{City} & \multirow{8}{*}{Population} & \multicolumn{2}{c}{\multirow{2}{*}{\parbox{1cm}{\centering Integration Grid}}} & \multicolumn{2}{c}{\multirow{2}{*}{\parbox{1cm}{\centering Traffic Control}}} & \multicolumn{2}{c}{\multirow{2}{*}{\parbox{1cm}{\centering Biking/Car Sharing}}}\\
\\
\cline{4-9}\\
& & & \rotatebox{90}{\centering Program} & \rotatebox{90}{\centering Data protection} & \rotatebox{90}{\centering Program} & \rotatebox{90}{\centering Data protection} & \rotatebox{90}{\centering Program} & \rotatebox{90}{\centering Data protection} \\
\cline{2-9}
& Zurich & 341,730 & Y & Y & Y & Y & Y & Y \\
& Geneva & 183,981 & Y & N/A & Y & N/A & Y & N/A \\
& Basel & 164,488 & Y & N/A & Y & N/A & Y & N/A \\
& Lausanne & 139,111 & Y & N/A & Y & N/A & Y & N/A \\
& Bern & 121,631 & Y & N/A & N/A & N/A & N/A & N/A \\
& Wintherthur & 91,908 & Y & N/A & Y & N/A & Y & N/A \\
& Luzern & 81,691 & Y & Y & Y & N/A & Y & N/A \\
& Lugano ** & 63,000 & Y & N/A & Y & N/A & Y & N/A \\
& Bellinzona ** & 16,572 & Y & N/A & Y & N/A & Y & N/A \\
\cline{2-9}
\\
\cline{2-9}
\multirow{20}{*}{\rotatebox{90}{Italy}} & & & \multicolumn{6}{c}{Smart Mobility Policies} \\
\cline{4-9}
& \multirow{8}{*}{City} & \multirow{8}{*}{Population} & \multicolumn{2}{c}{\multirow{2}{*}{\parbox{1cm}{\centering Integration Grid}}} & \multicolumn{2}{c}{\multirow{2}{*}{\parbox{1cm}{\centering Traffic Control}}} & \multicolumn{2}{c}{\multirow{2}{*}{\parbox{1cm}{\centering Biking/Car Sharing}}}\\
\\
\cline{4-9}\\
& & & \rotatebox{90}{\centering Program} & \rotatebox{90}{\centering Data protection} & \centering \rotatebox{90}{Program} & \rotatebox{90}{\centering Data protection} & \rotatebox{90}{\centering Program} & \rotatebox{90}{\centering Data protection} \\
\cline{2-9}
& Milan & 1,397,715 & Y & N/A & Y & Y & Y & N/A \\
& Turin & 848,196 & Y & N/A & Y & N/A & Y & N/A \\
& Genoa & 558,930 & Y & N/A & Y & N/A & Y & N/A \\
& Roma & 2,783,809 & Y & Y & Y & Y & Y & N/A \\
& Bologna & 394,463 & Y & N/A & Y & N/A & Y & N/A \\
& Florence & 359,755 & Y & N/A & Y & N/A & Y & N/A \\
& Naples & 940,940 & N/A & N/A & Y & N/A & Y & N/A \\
& Palermo & 640,720 & N/A & N/A & N/A & N/A & Y & N/A \\
& Bari & 313,003 & N/A & N/A & N/A & N/A & Y & N/A \\
\cline{2-9} \\
\end{tabular}
}
\caption{Selected smart city programs and corresponding data protection initiatives~\cite{Fabregue-B2023}}
\label{tab:data_scale}
\end{table}

The role of data exchange is crucial in MaaS, and there are different stakeholders who need to access and process data to function properly. Cottrill~\cite{Cottrill-C2020} emphasised the significance of examining third-party processors (e.g., payment processors, and hosting providers) of MaaS applications in terms of privacy considerations and implications. Pitera and Marinelli ~\cite{Pitera-K2017} and He and Chow~\cite{He-B2021} recommended to have agreements on the type and format of the data that can be shared among different actors of MaaS to best facilitate platform operations. When sharing data via an open data platform, a certain level of privacy control is needed to help public agencies better measure and evaluate the market~\cite{He-B2021}. Moreover, Butler et al.~\cite{Butler-L2021} indicated that one particular barrier to MaaS adoption is the privacy risk and concern related to access to personal data by nefarious sources. Butler et al.~\cite{Butler-L2021} also found out that the breach of intellectual properties would lead to businesses losing competition advantage. To this end, the importance of privacy regulations was highlighted in the study for the development of MaaS and to maintain the trust of both users and providers~\cite{Butler-L2021}. Along this line of research, researchers have examined whether real-world MaaS systems have policies/measures in place to address these issues. A study~\cite{Fabregue-B2023} investigated the development and deployment of data protection mechanisms in smart cities and MaaS, as illustrated in Table~\ref{tab:data_scale}, revealing poor data protection practices across different cities in Italy and Switzerland to address data privacy risks while deploying smart city programmes. To respond to this, regulations and policies, such as the EU's General Data Protection Regulation (GDPR), are anticipated to affect the deployment of MaaS significantly. The MaaS ecosystem involving both data controllers and processors must be designed and implemented to comply with these regulations. This includes adhering to principles such as privacy by design, explicit consent, and data protection as mandated by the GDPR~\cite{Cottrill-C2020}. However, to the best of our knowledge, research about MaaS is less studied to cover these perspectives.

\subsection{Data-Computer Pipeline Attacks}

Given the potential privacy concerns and risks mentioned above, researchers and practitioners have examined cyber attacks that could harm the MaaS ecosystem, including the illegal exploitation of sensitive data and disruptions to its availability and functionality. These threats can adversely affect individuals and society, impacting safety, security, and economic stability. To summarise, these associated attacks can be broadly classified into technical and socio-technical perspectives. We present these in more detail in the rest of this section.

\subsubsection{Technical Perspectives}

Denial-of-service (DoS) attacks have been widely recognised as a critical cyber security attack vector that every designer and developer of MaaS systems should be aware of, as suggested by several studies~\cite{Utriainen-R2018, Thai-J2018, Vaidya-B2020}. In a DoS attack, malicious actors exploit vulnerabilities in the system's infrastructure, overwhelming it with a flood of traffic or exhausting its resources to disrupt its normal operation, making it unavailable to legitimate customers. Thai et al.~\cite{Thai-J2018} suggested that the increasing popularity of MaaS could attract malicious parties' attention to launch DoS attacks to gain illicit advantages. To address this challenge, Thai et al.~\cite{Thai-J2018} introduced a theoretical framework aimed at mitigating the impact of DoS attacks on MaaS systems. Their framework involves analysing the network and considering it as a stochastic control problem with the main objectives of 1) maximising passenger retention within the network in a steady state; and 2) analysing the financial impacts of DoS attacks on MaaS systems. Apart from DoS attacks, several other attack types such as eavesdropping, spoofing attacks, jamming attacks, hijacking attacks, man-in-the-middle (MitM) attacks, replay attacks, relay attacks, remotely exploitable attacks, and ransomware attacks, have also been identified~\cite{Vaidya-B2020, Callegati-F2017a, Carvalho-G2019}.

Depending on how a MaaS system is developed, different approaches have been proposed to mitigate such attacks. For a MaaS system that integrates with edge-oriented computing (EOC), Carvalho et al.~\cite{Carvalho-G2019} suggested deploying machine learning based methodologies in edge servers to help detect EOC-related attacks and security breaches such as fault injection, user impersonation, crowd-sourcing attacks, and data manipulation. Furthermore, Nguyen et al.~\cite{Nguyen-T2019} proposed a blockchain-based approach for MaaS, aiming to enhance trust and transparency among stakeholders. Their proposal involves the utilisation of distributed computational resources allocated to various transport providers situated at the network's edge. Similarly, several studies~\cite{Costantini-F2019, Cruz-C2020} suggested that the utilisation of smart contracts on the blockchain could make payment and exchange of other services in MaaS systems somewhat faster and more convenient while also ensuring customer privacy when sharing data. Nguyen et al.~\cite{Nguyen-T2019} further suggested the need of clear specification on the contractual terms and statements, which can be achieved by applying cryptographic zero-knowledge argument schemes (SNARKs) to demonstrate that the terms can be satisfied and agreed on, without disclosing private information related to passengers in the contract. In another study, Bothos et.al.~\cite{Bothos-E2019} also addressed that the blockchain-based approach could allow conditional transactions by encrypting the data with selective access rights, which can enhance the security and privacy of personal information. The risks of these blockchain-based approaches can be quantified by the risk assessment framework~\cite{al2021cyber}.

\subsubsection{Socio-Technical Perspectives}

In addition to those technical threats given above, there are also some socio-technical threats that have been identified from past research. Cruz et.al.~\cite{Cruz-C2020} adopted a SWOT (Strength-Weakness-Opportunities-Threats) analysis for MaaS used in Lisbon, and identified a number of social issues as potential security threats: 1) reluctance of MaaS consumers to have control over their own apps; 2) conflicting objectives between private and public companies; and 3) unclear regulatory frameworks and data privacy issues. In another study, Vaidya and Mouftah~\cite{Vaidya-B2020} concluded that dishonest individuals, hackers, criminal groups, and dishonest organisations can pose potential security threats to MaaS systems. They also indicated that the main motivations of wrongdoings by dishonest individuals include financial gain and/or theft, leading to potential harms such as identity theft, vehicle theft, and financial loss for both individuals and organisations. Criminal groups have been identified as being motivated by a desire to cause service or business disruptions, leading to significant harm to individuals and organisations. For dishonest organisations, the main motives are to sabotage competitors, and/or industrial espionage~\cite{Vaidya-B2020}. The illegal and unethical use of personal data by MaaS operators/providers is another security risk recognised by some studies~\cite{Zhang-Z2021, Callegati-F2018}. According to the findings of Callegati et al.~\cite{Callegati-F2018}, malicious activities by MaaS operators pose potential security threats. These activities include the transmission of relevant information to competitors and the extraction of sensitive data by aggregating anonymised data. In addition, Callegati et.al.~\cite{Callegati-F2018} also suggested that the action of disabling the GPS device on drivers' vehicles can be considered as an insider threat, as it can potentially compromise the reliability of the GPS positioning system and other services that depend on it. Disabling geo-location data would undermine vehicle-to-grid services that transport vehicles and charging systems could offer in low-carbon MaaS systems, highlighting increasing trade-offs for mitigation of cyber security risks vs digitisation of energy and transport systems. Increasing inter-dependencies of transport and energy systems point to the importance of AI in monitoring, planning and coordinating the decisions of millions of individuals and assets like homes, transport vehicles, solar panels, and charging stations. Following an assessment of future challenges of privacy and data risks, we then discuss the risks associated with the use of AI algorithms in MaaS systems.

\subsection{Future Trends and Challenges}

In examining privacy and security risks for MaaS systems, we identified a notable gap in the existing research, which solely focuses on privacy risks and concerns, addressing two main areas of data handling: 1) the risks associated with profiling; and 2) the concerns related to third-party access. However, given the extensive data that MaaS can collect and process and the multiple parties involved, privacy risks and concerns could arise at various phases, including data generation, storage, processing, and sharing. We envisage that one of the future research directions is to conduct systematic research to comprehensively understand the full scope and depth of issues related to different phases of the MaaS data processing pipeline.

Moreover, privacy and security are often discussed together in the literature, yet important distinctions exist between them. Privacy is about the appropriate use and governance of personal data, ensuring it is collected, used, and shared properly. On the other hand, security focuses on protecting assets such as data and systems from malicious attacks and misuse~\cite{Jing-Q2014}. While security is essential for data protection, it may not be sufficient to ensure privacy~\cite{Jain-P2016}. We have introduced various security risks and attacks from both technical and socio-technical perspectives in this paper, however, considering the heterogeneity of MaaS, in-depth discussions on the relationship between privacy and security are necessary to design better MaaS systems. In addition, how to balance benefits with safety, data security, privacy, equity, and market distortion~\cite{Enoch-M2023} remains another challenge to designing MaaS systems that can be commercially and socially sustainable.

Furthermore, the implications of introducing and enforcing regulations like the GDPR on MaaS are under-explored, particularly in terms of their impact and the measures stakeholders can take to comply with these regulations to enhance the security of the MaaS ecosystem and privacy of their consumers. These remain open questions and future challenges. We would encourage researchers, practitioners, and policymakers to collaborate in developing comprehensive guidance and requirements for MaaS from such a perspective.

\section{Review of Attacks on AI Algorithms and Defence Measures}
\label{section:review_AI}

Current and emerging adversarial AI attacks (evasion, extraction, gamification) are reviewed in the context of the MaaS ecosystem. These novel attacks often combine novel methods (e.g., inverse learning) with traditional attack vectors (e.g., man-in-the-middle) reviewed previously.

\subsection{AI Surfaces for Attack}

The personal data collected from MaaS consumers can enable personalised optimisation and adaptation to time-varying consumer behaviours, where past experiences affect future requirements and satisfaction. This is why Deep Reinforcement Learning (DRL) algorithms are appealing compared to one-shot heuristic optimisation based on instantaneous preferences, because every resource allocation decision will affect not only current performance but also future consumer preferences in a causal way. For example, we may overtime wish to balance the fairness of transport provisioning with the consumer experience, both of which are dynamic attributes and have memory properties (e.g., a MaaS consumer's likelihood of future use is based on their past experiences)~\cite{chu2023deep}. As these likelihoods are hidden and dynamic, we cannot do instant one-shot optimisation or try to guess a general static behaviour function.

\begin{figure}[!htb]
\centering
\includegraphics[width=\linewidth]{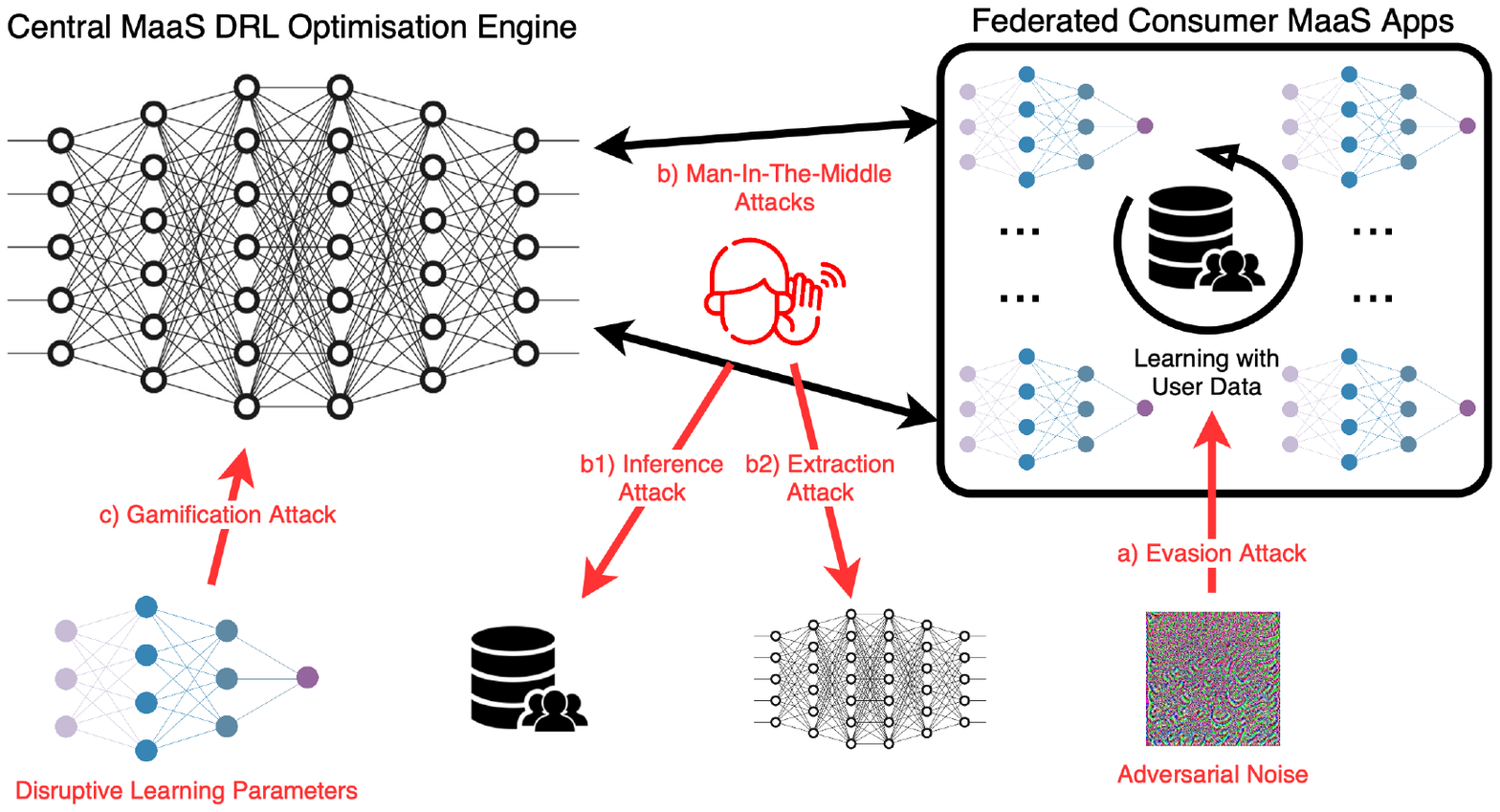}
\caption{Attack vectors against a machine learning MaaS controller: a) evasion data attacks that attempt to cause mis-performance in the machine learning algorithm, b) MitM attacks, b1) inferring personal information by eavesdropping on hyper-parameter or raw data exchanges, b2) model extraction attacks try to infer the overall MaaS controller model, and d) gamification attacks that use a sequence of false data to erode system wide performance.}
\label{fig:AI_attack_vectors}
\end{figure}

Whilst DRL advance over conventional hidden Markov model (HMM) approaches, the integrity and confidentiality of DRL models and its data are threatened by data leakage, data inference/estimation, and direct cyber attacks~\cite{zou2021cyber}. This can create serious issues for violating personal privacy in terms of transport choices and behaviours, as well as national safety in terms of disrupting transport and energy provision (in the case of low-carbon MaaS). Whilst federated learning (FL) (see Fig.~\ref{fig:AI_attack_vectors}) can be a promising approach to address certain data privacy issues for MaaS~\cite{chu2023federated, chu2024privacy}, there remain challenges that we discuss below.

\subsection{Algorithm Attacks}

\subsubsection{Evasion Attacks}

Evasion attacks are data manipulation attacks targeting particular types of layers, such as convolution layers, as neural network architectures typically consist of a variety of layers to perform tasks such as regression, classification, and text generation. Consequently, any deep learning architecture that relies on convolution tasks will suffer to varying degrees from evasion attacks. Here, we review specific recent developments in evasion attacks relevant to MaaS, as illustrated in Fig.~\ref{fig:AI_attack_vectors}a. Evasion attacks can degrade performance in different layers of neural networks for both centralised and federated learning approaches, resulting in errors or insensitivity to consumer demands.

The majority of evasion attacks focus on computer vision applications, which, in MaaS scenarios, translate to real-world perception of intelligent transport, e.g., self-driving vehicles and intelligent traffic infrastructures. In 2019, Thys et al.~\cite{foolyolov2} proposed a basic patch method to fool a YOLO v2 detector, demonstrating CNN detection evasion. This method was extended to the context of aerial surveillance photography for a YOLO v2 object detector, but performance was analysed based on flat images without noise, transformation, or discoloration. In 2020, Xu et al.~\cite{mitTshirt} designed practical evasion attacks, considering both the deformation of evasion patterns caused by movement and the data spectrum differences between the desired and achievable evasion attacks. Consequently, more recent work has focused on training evasion attack data patterns constrained by reality. Of particular relevance to MaaS, several novel variants of patch training using GANs~\cite{GANpatch} have been developed: 1) generating evasion data patches with data dimensions constrained to personal requirements~\cite{bugpatch} (e.g., different consumers may have varying data on requirements and preferences); and 2) enabling data dimensions to be flexible to transformations and deformations~\cite{multi-shape, rotation_sensitive, softbody} (e.g., consumers may dramatically change their personal attributes or travel requirements). These constraints are embedded in the evasion pattern generation process by adding general boundary constraints or transformations to the data to mimic real-world scenarios.

Evasion attacks on Large Language Models (LLMs) also pose significant risks. Zou et al.~\cite{zou2023universal} proposed an approach that generates malicious inputs which are efficient and transferable. These attacks manipulate the input data to evade detection mechanisms and compromise the integrity of LLM-based applications.

\subsubsection{Membership Inference and Model Extraction Attacks}

Membership Inference Attacks (MIAs) and Model Extraction Attacks (MEAs) also represent formidable security challenges to MaaS, as illustrated in Fig.~\ref{fig:AI_attack_vectors}b. The former can disclose whether specific user data or travel patterns have been utilised in training MaaS algorithms, leading to privacy breaches, while the latter might replicate proprietary algorithms, eroding competitive edges. Although Federated Learning (FL), a distributed ML paradigm that keeps private data on client devices for local training and then aggregates these models on a server, is widely adopted in MaaS and lauded for enhancing privacy, it is not impervious to these threats. The server-client architecture of FL increases the opportunity for adversaries to perform Man-in-the-Middle (MitM) attacks, threatening AI algorithms via communication channels and household client devices, as shown in Fig.~\ref{fig:AI_attack_vectors}. MIAs (b1) can detect data patterns from model updates, while MEAs (b2) can directly target the aggregated model in the client model update process.

\begin{itemize}
 \item \textbf{Member Inference Attacks (MIAs):} Such attacks were first described by Shokri et al.~\cite{ShokriMIA}, aiming to infer whether a data record is a member of the training set of the target model. They achieved this by training an attack model with the output vectors of the target model, reporting over 90\% and 70\% inference accuracy on Google and Amazon services, respectively. Nasr et al.~\cite{NasrMIAFL} presented a comprehensive analysis of the threats of MIAs on FL. Their experiments compared attacks from both categories of participants, an adversarial aggregator and malicious clients, with different levels of access to the target model. A malicious client can achieve 76.7\% inference accuracy, and a malicious server can achieve 82.1\%. Hu et al.~\cite{husourceinferattack} extended such attacks to infer the source of data records, i.e., the specific client that owns the identified data records.

 \item \textbf{Model Extraction Attacks (MEAs):} MEAs can be seen as a malicious application of the techniques described by Hinton et al.~\cite{hinton2015distilling}, targeting Machine Learning as a Service (MLaaS). MaaS with an FL architecture naturally inherits susceptibility to MEAs. Tramer et al.~\cite{tramer2016stealing} first introduced MEAs that train a replica model through prediction queries to the target model. Orekondy et al.~\cite{OrekondyKnockoffNets} proposed Knockoff Nets, which employ reinforcement learning to efficiently select samples for queries, producing a similarly performant replica model of a complex model with a compact architecture at a query cost as low as \$30.
\end{itemize} 

In addition to privacy violations and intellectual property breaches, both MIAs and MEAs can serve as stepping stones for evasion attacks.

\subsubsection{Gamification Attacks}

DRL is an effective method for solving problems modelled as Markov decision processes (MDPs) by approximating the optimal action-value function or policy. In the context of MaaS, many operational tasks can be formalised as MDPs and thus they can be addressed using DRL. For example, the problem of multi-modal journey planning can be represented as a Markov model where consumer experiences are characterised by a transient effect on future satisfaction and retention~\cite{chu2023deep}, making the MDP state a representation of the consumer and traffic status, such as consumer profile, satisfaction with the service, and traffic congestion. Although the input of consumer status provides sufficient information for accommodating heterogeneous consumers with dynamic preferences, it creates a channel for consumers to deceive the system by providing false information for their own benefit. During both the training and inference stages, the DRL agent relies on the status of the consumer to perform the model inference. Therefore, incorrect states can influence the learning process of the state transition model of the environment during the training stage, as well as lead to inappropriate output actions during the inference stage. This scenario creates a game between the MaaS planner and consumers with different utility functions (see Fig.~\ref{fig:AI_attack_vectors}d).

In this gamification attack, the planner aims to maximise the total utility function, such as consumer satisfaction, in order to increase profit through higher consumer retention. On the other hand, malicious consumers with selfish or destructive utility functions provide false profiles and experiences to the planner. Recently, a new reinforcement learning-based attack called the passenger spoofing attack was identified~\cite{chu2023passenger, chu2024multi}, which reduces the profit and consumer satisfaction of the MaaS system. Another approach is inverse-learning, where an attack learns the reward or objective function of the system (often with the help of some prior model knowledge)~\cite{PERRUSQUIA2023396}. Knowledge of this can be used to gamify the system. In either case, subsequently a malicious consumer spoofs as a consumer with the same origin and destination as another regular consumer. The main difference between these two types of consumers is that the malicious consumer is providing a fake profile and satisfaction. The fake profile and satisfaction are generated by a reinforcement learning-based malicious agent to increase the probability of being prioritised by the MaaS planner over the regular consumer for the malicious goal, such as allocating the limited and ideal mobility service to the malicious consumer instead of the regular consumer. Additionally, these malicious consumers can team up to strengthen the attack. The study also discovered that the attack can be strengthened by multi-agent reinforcement learning, which considers the spatial distribution among the malicious agents and consumers. Consequently, the travel time, cost, and satisfaction of regular consumers may be degraded due to the attack, and thus affecting the consumer experience and retainment.

Behzadan and Munir~\cite{behzadan2017vulnerability} conducted a study to investigate the vulnerability of Deep Q-networks (DQNs) and to evaluate the transferrability of adversarial examples across different DQN models. They devised an attack that involves perturbing the states from the environment, leading the DRL agent to execute adversary-desired actions based on the perturbed states. The attack comprises two phases: initialisation and exploitation. In the initialisation phase, an adversarial policy is first obtained by training a DQN based on an adversarial reward function. Then, a replica of the target's DQN is created and initialised from random parameters. In the exploitation phase, adversarial inputs are crafted using adversarial example crafting techniques such as the Fast Gradient Sign Method~\cite{goodfellow2015explaining} and Jacobian-based Saliency Map Attack~\cite{papernot2016limitations}, and their amplitude is controlled to ensure the perturbations are imperceptible. These crafted states cause the target DQN to follow actions determined by the adversarial policy. Consequently, the attacker can manipulate the DQN's learning process and leading to incorrect optimal action choices. Furthermore, the authors manipulated the policy of the DQN by exploiting the transferability of adversarial samples. They used a black-box setting to demonstrate the success rate of their method, achieving a success rate of 70\% when adversarial examples were transferred from one model to another.

\subsection{Mitigation and Countermeasures}

\subsubsection{Evasion Attacks}

In principle, defence measures against evasion attacks mainly aim to improve the robustness of the model itself in design and training process. Common approaches for computer vision tasks include the following.

\begin{itemize}
\item Adversarial training: Adversarial training, adding adversarial examples to the training set, was proven to be effective in weakening the attack of adversarial samples of the same origin and level of perturbation~\cite{goodfellow2015explaining}. Yet, the effectiveness declines significantly when applied against adversarial examples generated by another model. There are also alternative approaches for adversarial training. Metzen et al.~\cite{metzendetectadversary} proposed a sub-network design, which is a binary classifier within the network that discriminate genuine data from perturbed data. Samangouei et al.~\cite{samangouei2018defense} proposed a generative adversarial network (GAN) that generates an unperturbed version of the given input, which is then fed into the classifier/detector. Empirical evidence suggested that these methods only take good effect when the source of attack is known, which makes it difficult to validate actual defence performance in experiments.

\item Region-based classifier: Cao et a.~\cite{CaoRegionBased} proposed region-based classification, sampling points in the immediate neighbouring space of the given image to serve as the input for the model. Such a method, equivalent to adding random noises to input, is geometrically intuitive for defending against adversarial examples of minor perturbations. However, when dealing with geometrical attacks such as DeepFool~\cite{Moosavi-DeepFool}, which takes the minimal distance vector to the decision boundary as the perturbation, region-based methods are impotent.

\item Defensive distillation: This~\cite{papernot2016distillation} is essentially the same technique as in knowledge distillation~\cite{hinton2015distilling}. By learning the soft labels provided by the teacher model, the decision boundary of the student model is smoothed, hence the higher difficulty for minor perturbations to push a sample cross boundary. Yet, unlike in knowledge distillation, in defensive distillation, the student model is not of a smaller capacity, but of the same as the teacher model. In addition, in defensive distillation, a temperature factor $T$ was added to the softmax function, as shown in the equation below, which controls the `hardness' of the maximum with negative correlation, i.e., when $T$ goes to 0, the function produces a quasi one-hot result, and when $T$ goes to infinity, the result is a uniform distribution. In practice, $T$ can be a large value in training but set to 1 in testing, which equivalently makes the inputs to the softmax function larger by the factor of $T$, consequently magnifying the difference of target class and the rest. As reported in~\cite{carlini2017towards}, the attack success rate dropped from 91\% to 0.5\%, when $T$ increased from 1 to 100.
\end{itemize}

As discussed above, for computer vision tasks, different defence approaches against evasion attack varies in advantageous scenarios, in MaaS systems, a combination of these defensive designs can largely weaken the threat posed by evasion attacks.

As for LLMs, the most common defence against adversarial prompting is safety alignment, i.e., aligning the LLM's behaviour with human values and norms. Approaches of safety alignment mainly take place in the training process and output generation. Mechanisms such as reward engineering, reinforcement learning from human feedback, LLMs-as-a-Judge can be applied to the training process, and rule-based constraints or other output filtering or moderation methods can be applied to the output generation process.

\subsubsection{Membership Inference Attacks and Model Extraction Attacks}

MIAs and MEAs take similar approaches to threat AI-enabled MaaS systems, through APIs, FL interactions, or communication channels, but with different purposes of violating privacy and stealing proprietary models. Defence methodologies naturally align with their respective malicious intention.

\begin{itemize}
\item Member Inference Attacks (MIAs): Differential Privacy (DP) is the most straightforward defence methods against MIAs. Abadi et al.~\cite{Abadi2016Deep} proposed Differentially Private Stochastic Gradient Descent (DP-SGD), first introducing differential privacy to deep learning applications. They first clip gradients, then add random noise to the clipped gradients, where both process are parameter. Such a method involve a trade-off between privacy preservation and model performance. Nasr et al.~\cite{Nasr2018Machine} proposed Adversarial Regularisation to optimise the trade-off, maximising the joint score of privacy preservation and model accuracy. Yu et al.~\cite{Yu2021GEP} proposed Gradient Embedding Perturbation (GEP), in line with other DP approaches, which projects private gradients into a non-sensitive anchor space to produce a low dimensional gradient embedding, which is perturbed according to the privacy budget. They reported superior performance with reasonable computational cost and modest privacy guarantee. Tang et al.~\cite{Tang2022Mitigating} introduced the SELENA framework, which address the defence with ensemble training multiple models with subsets of the training set and self distillation, outperforming Adversarial Regularisation techniques.

\item Model Extraction Attacks (MEAs): Such attacks exploit the output of the target model to create one or more unauthorised replicas. A primary strategy for defending against these attacks involves degrading the performance of the extracted or cloned models. Current defence methodologies fall into two main categories: output perturbation and malicious query detection. Output perturbation techniques include quantisation of output conﬁdence scores~\cite{tramer2016stealing}, concealment of sub-optimal predictions~\cite{OrekondyKnockoffNets}, injection of uncertainty towards the end of the posterior distribution~\cite{Lee2019Defending}. The efficacy of these defences is contingent upon the degree of perturbation implemented. Aligning with these methods, Orekondy et al.~\cite{Orekondy2020Prediction} proposed a proactive defence mechanism by poisoning the posterior probabilities. The injected perturbation is designed to misdirect the optimisation of the attacker's replica model. On the detection front, Juuti et al.~\cite{Juuti2019PRADA} proposed a method for detecting malicious queries. However, their approach is predicated on two stringent and potentially unrealistic assumptions that the attacker's query samples are closely distributed, and the attacker only queries as one user instead of multiple users with smaller query batches. Kariyappa et al.~\cite{Kariyappa2020AM} proposed Adaptive Misinformation, which is a combination of output perturbation and malicious query detection, but emphasised on Out-Of-Distribution (OOD) inputs. Their detector is trained to identify OOD inputs from In-Distribution (ID) input, assuming adversarial queries using the former, then inject misinformation to the outputs of malicious queries. They also proposed Ensemble of Diverse Models (EDM)~\cite{Kariyappa2021EDM}, to defend against model extraction attacks by training multiple different models and use a random one for each query.
\end{itemize}

It is worth noting that actual adversaries can couple MIAs and MEAs, as the two can potentially reinforce each other. Hence, it is important to put joint attacks into consideration when designing defence strategies. Yet, how these joint defence strategies may be implemented across different actors in MaaS ecosystem and their impacts on business models are overlooked as we discussed in the next section.

\subsubsection{Gamification Attacks}

Mitigation methods for gamification attacks are still under-explored. On one hand, adversarial training mentioned above could also improve the robustness of deep reinforcement learning models such as DQNs. While on the other, as previously mentioned, this attack can benefit from volumne, e.g., attackers with more malicious clients are more successful. Therefore, when designing defence mechanisms, it it important to balance the robustness gain and defence overhead, avoiding potential arms race with attackers.

\subsection{Future Trends and Challenges}

Research in adversarial attacks and defences is anticipated to remain highly active. While current attacks are mainly white-box attacks or grey-box attacks, i.e., assuming complete or partial knowledge of the victim model, are expected to further develop, innovative techniques for black-box attacks, particularly query-based ones, will likely become more prominent. Also, attacks on transformer-based models in computer vision and both natural language processing (NLP), which are gaining popularity, are also expected.

On the defence side, developing intrinsically robust models, managing the robustness-accuracy trade-off, adversarial training, and certified defences will continue to be major research focuses. 

Adversarial attacks are also expected to expand into multi-modal tasks, where multiple types of input, e.g., text, images, audio and video, are fed to the model. Tackling the challenge posed by adversarial attacks, such as evasion, extraction and inference attacks, gamification attacks, to the reliability of AI-equipped systems will remain vital.

\section{Impact on MaaS Business Ecosystem}
\label{business_section}

While MaaS scholars have recognised how the decarbonisation of transport may impact the MaaS ecosystem, its implications on cyber security and privacy concerns have been largely ignored so far with the exception of Alderete-Peralta and Balta-Ozkan~\cite{Alderete2024road}. In this section, we discuss the impacts of AI-related risks on MaaS business models. Firstly, we argue that the future MaaS ecosystem is not limited to electric powertrains and that may incorporate hydrogen ones as well. The electric MaaS (eMaaS) is proposed to compose a combination of MaaS, electric mobility systems and shared electric mobility systems~\cite{reyes2020state}. Given the expected role of hydrogen fuel cell technologies in decarbonisation of transport systems, nationally or globally~\cite{Iea, park2022much}, we define low-carbon MaaS (locMaaS) to include both hydrogen and electric powertrains as well as relevant physical infrastructure and services (e.g., battery management). While the pace, nature, and extent of the roles of hydrogen fuel cell and electric powertrain technologies to decarbonise road and rail transport are likely to vary across different countries, the business model and ecosystem of locMaaS will definitely need to rely on digital data, technologies and services.

Secondly, digitisation will have profound impacts on locMaaS ecosystem and business models. Given different strategies Original Equipment Manufacturers (OEMs) are adopting to collect and control multi-dimensional data from the users as well as battery and charging network conditions (e.g., Volvo collaborating with a digital service product and service developer, Siemens), researchers have noted a more distributed decision-making structure among OEM value chain actors~\cite{perez2023mobility}. Yet, who these actors are and how their roles may evolve to manage cyber security risks have not been sufficiently understood. The eMaaS ecosystem proposed by Garcia et al.~\cite{reyes2020state} divides the actors in terms of core business, extended enterprise and business ecosystem but it neglects the role of operators of physical energy infrastructure and energy market actors. A more recent study by Anthony~\cite{anthony2023data} discusses the importance of energy infrastructure but it overlooks where both energy infrastructure and market actors sit within the eMaaS ecosystem that is limited to the business models, parking models, pricing models, and payment models. We argue that locMaaS business models in the future will need to develop wider and deeper collaborations across OEMs, physical energy infrastructure operators (e.g., charging station operators, distribution and transmission network operators, hydrogen fuelling systems) and energy markets. With increasing utilisation of renewable energy resources, managing power grids in terms of predictability of load, voltage and demand flows will require smart solutions. By controlling when, how much and where to charge, electric and hydrogen powertrains can provide vehicle-to-grid (V2G), vehicle-to-everything (V2X) and storage services to energy balancing and ancillary services markets which can be captured via suppliers, aggregators or vehicle owners directly, depending on the design and operation of energy markets as well as the functionalities of the vehicles themselves. Using the DRL or FL algorithms, locMaaS providers can reduce their (or transport operators') energy costs and emissions as well as generate additional income from taking part in energy balancing and ancillary services markets, at local or national levels by controlling the routing, and location, timing and state-of-charge of batteries. On the other hand, increasing digitisation of powertrain systems and other transport vehicles like scooters, and motorbikes point to data privacy risks extending beyond the users of locMaaS to include vehicles and devices with implications for the cyber security of energy networks. While AI offers great potential to observe and control the operation of many assets operating at the intersection of transport and energy networks, the downside is that there are more vulnerabilities that adversaries may try to exploit in the form of MIA or MEAs. How adversarial vulnerabilities may cascade and what action by which actor may lead to a more secure outcome across different actors in a MaaS business ecosystem are important questions that require further research.

\section{Conclusion}
\label{conclusion}

Future trends in transport algorithms are expected to leverage inherent characteristics of road networks, such as hierarchical structure, to enhance efficiency, while established techniques will likely maintain dominance due to their superior efficacy. The integration of real-world data in experimental validation processes will remain crucial for ensuring algorithmic fidelity and challenging conventional assumptions. Despite these advancements, persistent challenges include realising a global multimodal journey planner that efficiently incorporates real-time data and personalised elements, necessitating ongoing innovations. Moreover, the increasing reliance on data-driven and AI-based technologies in transport underscores the importance of robust cyber security strategies to safeguard user privacy and system integrity. In MaaS systems, addressing the full scope of privacy and security risks, particularly across different data processing phases, remains essential. This includes distinguishing between privacy, which governs appropriate data use, and security, which protects against malicious attacks. Systematic research and regulatory compliance, such as with the GDPR, are needed to enhance the MaaS ecosystem's security. Additionally, the field of adversarial attacks and defences will continue to evolve, with a focus on innovative black-box attack techniques, robustness in transformer-based models, and multi-modal tasks. Developing robust models and effective defences against diverse adversarial attacks will be critical to maintaining the reliability of AI-equipped systems.

This survey paper has provided a comprehensive overview of the intersection between AI-driven MaaS design and the various cyber security challenges it faces. As MaaS continues to integrate different transport modalities and relies on AI algorithms for personalised journey planning and optimisation, it becomes increasingly susceptible to a wide range of cyber attacks and privacy risks.

We have examined the diverse cyber security challenges facing MaaS, including privacy risks such as profiling, inference, and third-party threats, as well as adversarial AI attacks like evasion, extraction, and gamification. These risks pose significant threats to the integrity, confidentiality, and availability of MaaS systems, impacting both individual users and the broader ecosystem.

Moreover, the paper highlights the evolving nature of these risks, with attackers combining novel techniques with traditional attack vectors to exploit vulnerabilities in MaaS systems. From evasion attacks targeting AI algorithms to model extraction attacks compromising proprietary models, the breadth and sophistication of threats require robust countermeasures.

To mitigate these risks, MaaS stakeholders must prioritise cyber security measures at both the data and AI levels. This includes implementing defence mechanisms such as adversarial training, differential privacy, and output perturbation to bolster the resilience of AI algorithms against attacks.

Furthermore, collaboration across transport and energy sectors, including OEMs, transport and energy infrastructure operators, and energy markets, will be crucial in developing comprehensive and resilient business models for low-carbon MaaS. By leveraging advanced AI technologies and prioritising cyber security, MaaS can realise its potential to revolutionise transportation while safeguarding user privacy and system integrity.

\textbf{Contribution Statements:} KFC and WG contributed to design and writing of paper. JY, KFC, and WG wrote the adversarial AI part of paper. HY and SL contributed to the data privacy part of the paper, and NO wrote the business impact part of the paper. All authors proofread the final version of the paper to ensure its overall quality and smooth connection of different parts.

\bibliographystyle{IEEEtran}
\bibliography{references}

\vskip 0pt plus -1fil

\begin{IEEEbiography}[{\includegraphics[width=1in,height=1.25in,clip,keepaspectratio]{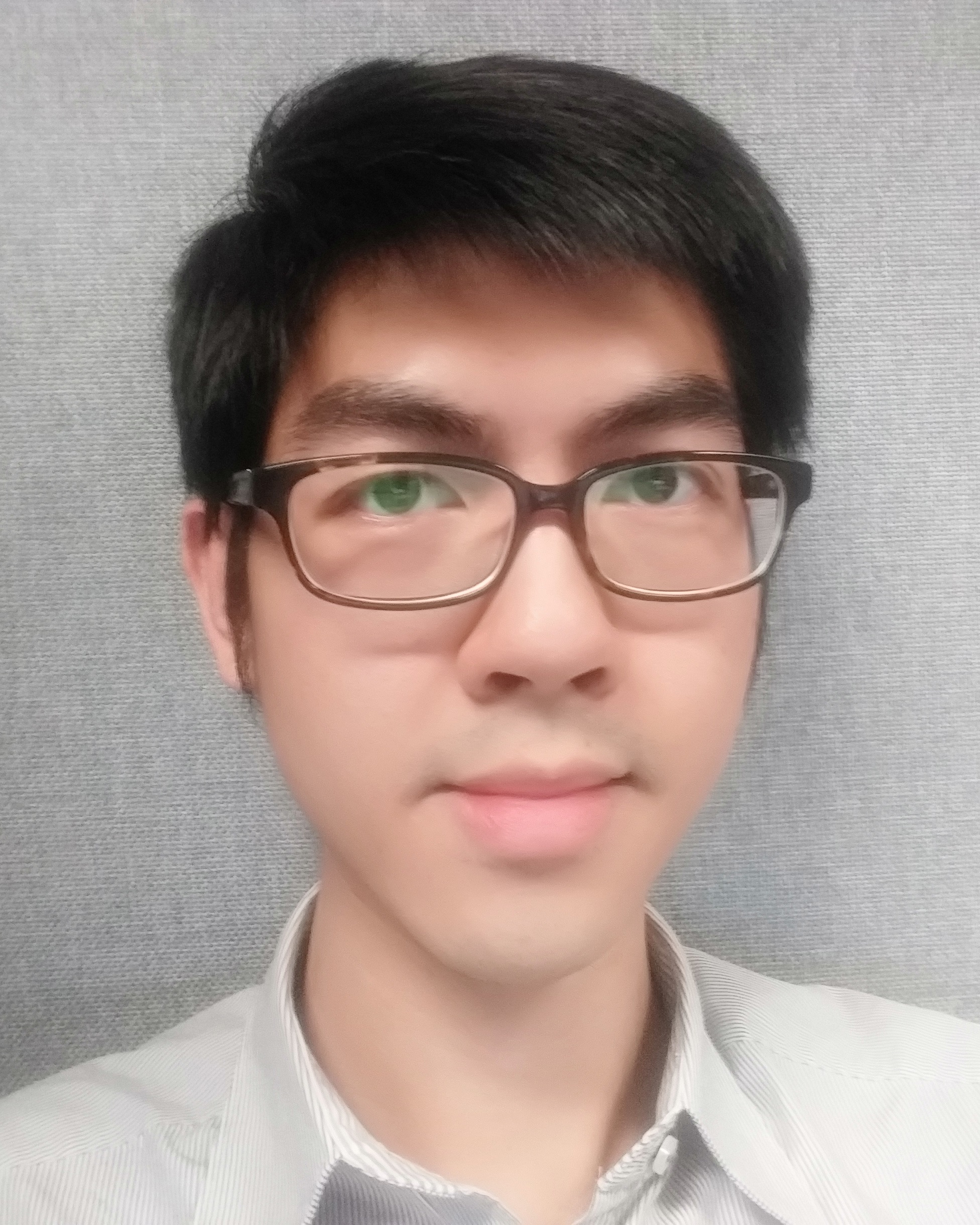}}]
{Kai-Fung Chu}
(Member, IEEE)
received the Ph.D.\ degree in Electrical and Electronic Engineering, The University of Hong Kong, Hong Kong, in 2020, and the M.Sc.\ and B.Eng.\ (First Class Honors) degrees both in Electronic and Information Engineering from The Hong Kong Polytechnic University, Hong Kong, in 2016 and 2013, respectively. He is a Marie Skłodowska-Curie Fellow at the University of Cambridge. He was a Research Assistant Professor at the Hong Kong Polytechnic University and a Research Fellow at Cranfield University. He also worked in the industry as an engineer for several years. His research interests include machine intelligence, embodied artificial intelligence, autonomous vehicles, and intelligent transportation systems.
\end{IEEEbiography}

\vskip 0pt plus -1fil

\begin{IEEEbiography}[{\includegraphics[width=1in,height=1.25in,clip,keepaspectratio]{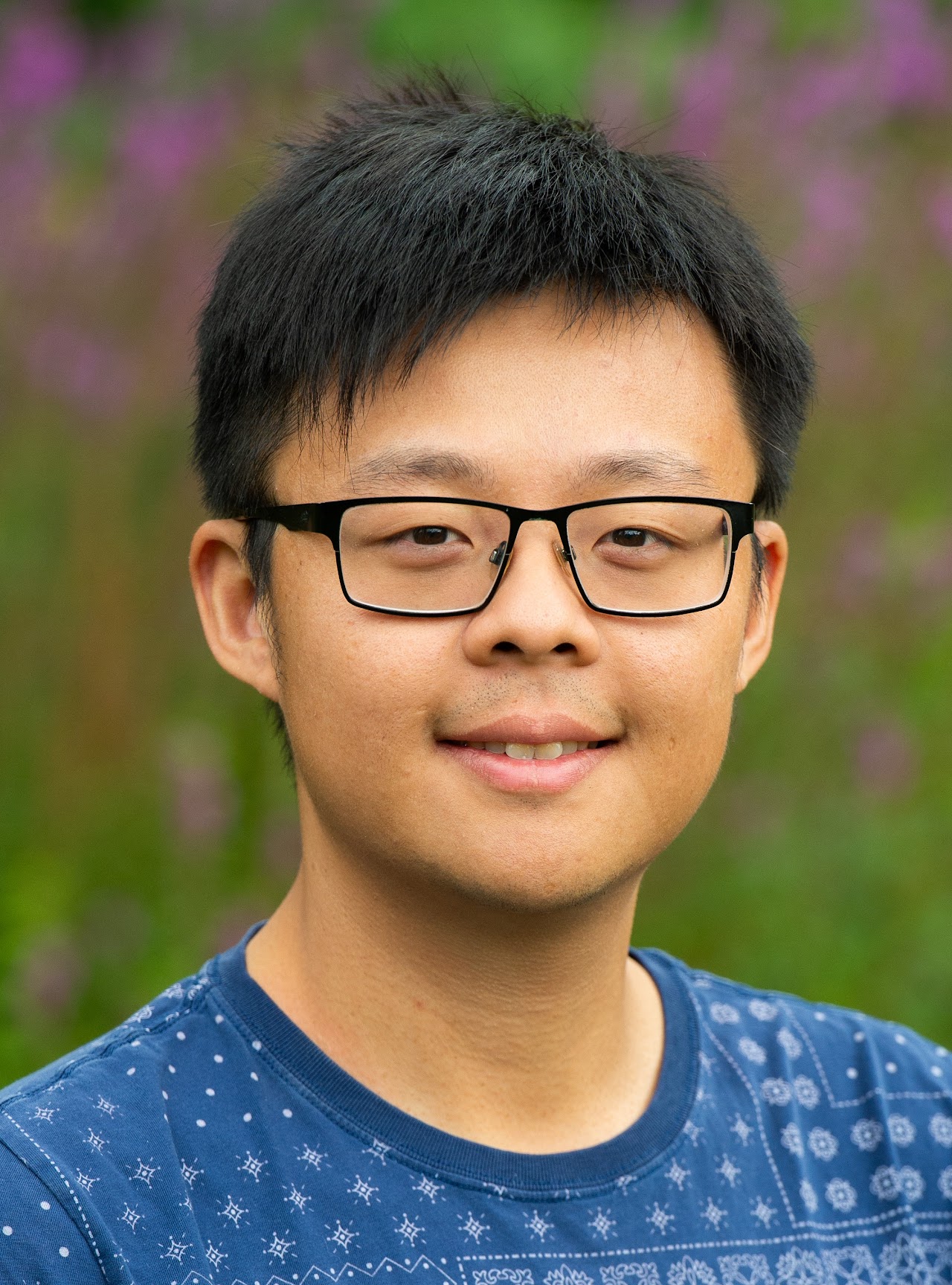}}]
{Haiyue Yuan}
(Member, IEEE)
received his BEng degree in Mobile Communication System in 2007 and MSc degree in Finance in 2009, both from University of Sheffield. He received his PhD degree in Electronic Engineering in 2013 from Centre for Vision, Speech and Signal Processing (CVSSP), University of Surrey. He is now a Research Associate, working at Institute of Cyber Security for Society (iCSS), University of Kent. He worked on a number of multidisciplinary projects at Department of Computer Science and CVSSP, University of Surrey. His main research interest include human computer interaction, and its intersection with cyber security and AI.
\end{IEEEbiography}

\vskip 0pt plus -1fil

\begin{IEEEbiography}[{\includegraphics[width=1in,height=1.25in,clip,keepaspectratio]{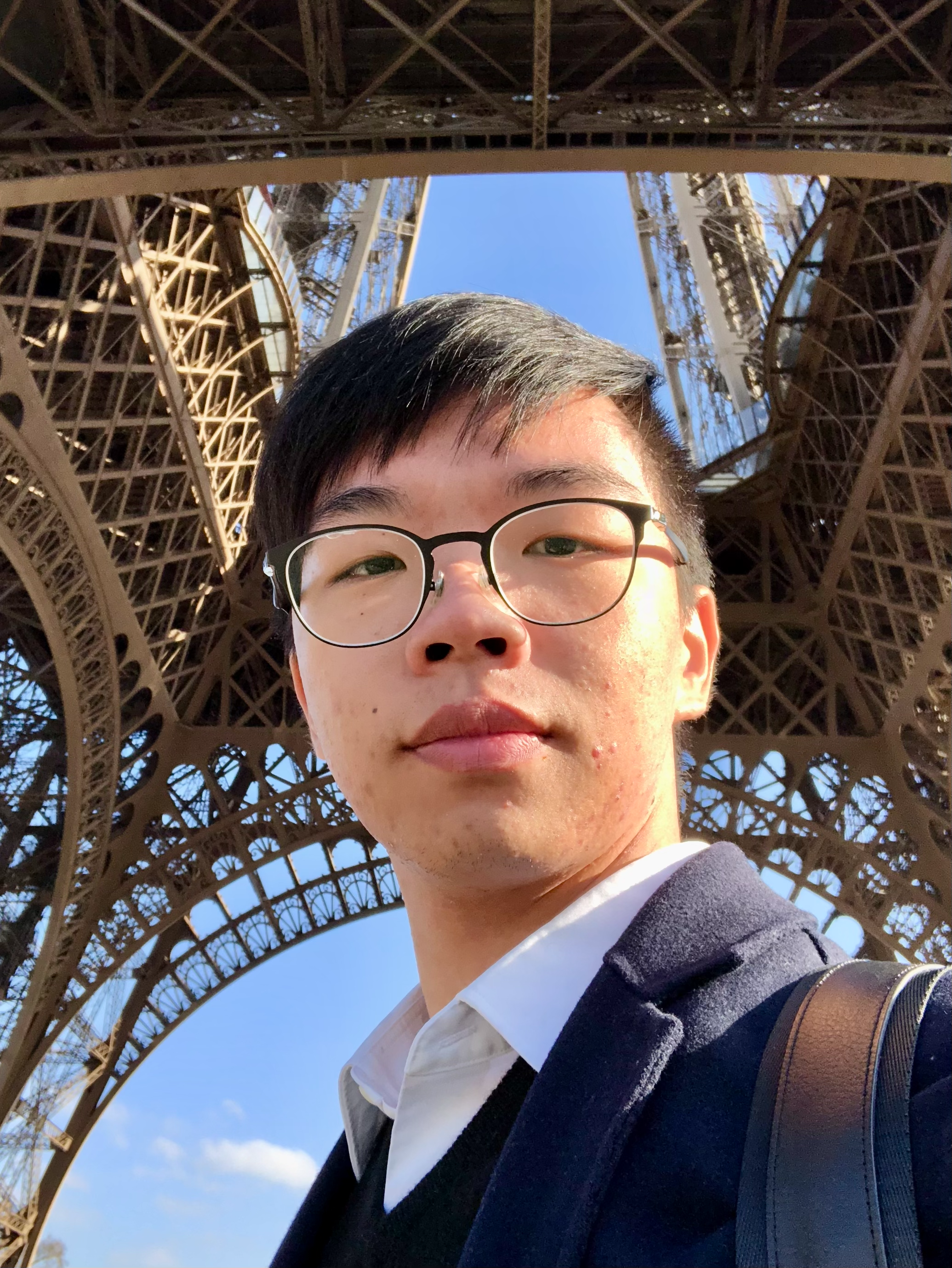}}]
{Jinsheng Yuan}
is currently a student pursuing  Ph.D.\ in Aerospace, at Cranfield University, UK. He received a M.Sc.\ in Computer Vision, Robotics and Machine Learning from from Centre for Vision, Speech and Signal Processing (CVSSP), University of Surrey, UK, in 2018 and B.Eng in Automation from South China University of Technology, China, in 2017. He also worked in the autonomous driving industry as an engineer developing camera perception algorithm. His research interests include AI safety, adversarial AI, approximate computing, topological data analysis.
\end{IEEEbiography}

\vskip 0pt plus -1fil

\begin{IEEEbiography}[{\includegraphics[width=1in,height=1.25in,clip,keepaspectratio]{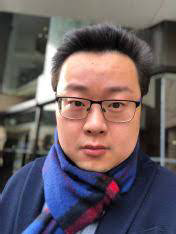}}]
{Weisi Guo}
(Senior Member, IEEE)
received his PhD degree from the University of Cambridge, UK in 2011. He was an Associate Professor at the University of Warwick (2012-19) and Turing Fellow (2017-19). He is currently a full Professor and head of the Human Machine Intelligence Group at Cranfield University, UK. He was a winner of the IET Innovation Award. His research interests focus on networks, data science, and autonomy.
\end{IEEEbiography}

\vskip 0pt plus -1fil

\begin{IEEEbiography}[{\includegraphics[width=1in,height=1.25in,clip,keepaspectratio]{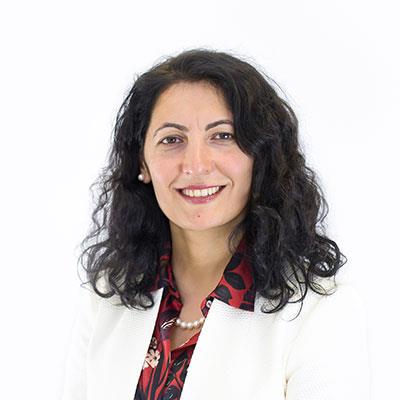}}]
{Nazmiye Balta-Ozkan}
(Member, IEEE)
received the BSc and MSc degrees in Urban and Regional Planning from Istanbul Technical University in 1995 and 1997, respectively, and the PhD in Regional Planning from University of Illinois Urbana-Champaign in 2004. She is a Professor in Sustainable Energy Transitions at Cranfield University. Her research focuses on developing tools, methods and frameworks for socio-technical construction of smart and low carbon energy and mobility systems.
\end{IEEEbiography}

\vskip 0pt plus -1fil

\begin{IEEEbiography}[{\includegraphics[width=1in,height=1.25in,clip,keepaspectratio]{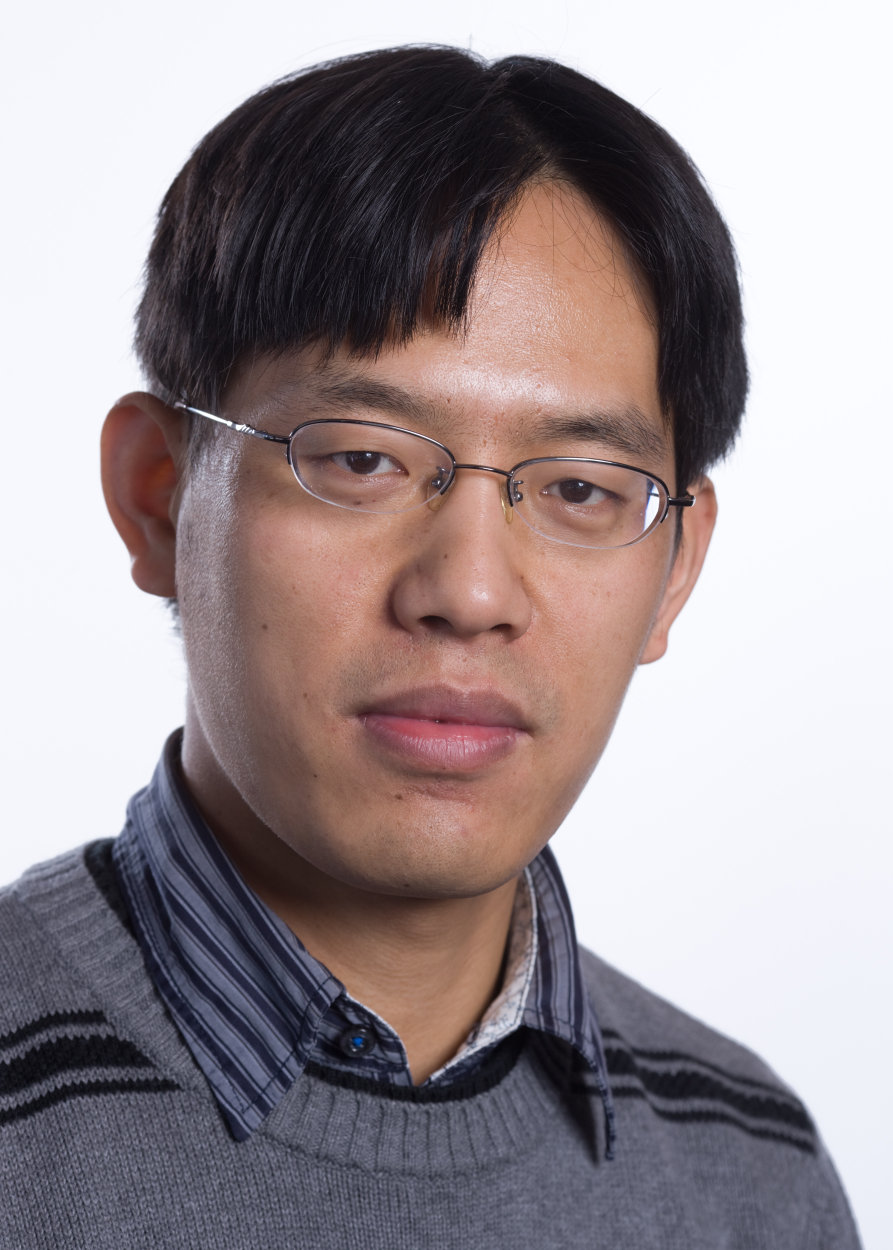}}]
{Shujun Li}
(Senior Member, IEEE)
received the BE degree in Information Science and Engineering, and the PhD degree in Information and Communication Engineering from Xi'an Jiaotong University, China, in 1997 and 2003, respectively. From 2003 to 2007, he was a postdoctoral research assistant with the City University of Hong Kong, and then a postdoctoral fellow with the Hong Kong Polytechnic University. From 2007 to 2008, he was conducting visiting research at FernUniversität, Hagen, Germany, as a Humboldt Research Fellow. From 2008 to 2011, he was a Zukunftskolleg Fellow at Universität Konstanz, Germany. In 2011, he joined the University of Surrey, UK, initially as a Senior Lecturer and then a Reader. Since 2017, he has been a Professor of Cyber Security with the University of Kent, UK, and currently directing the Institute of Cyber Security for Society (iCSS). His current research interests mainly focus on interplays between several interdisciplinary research areas, including cyber security and privacy, cybercrime and digital forensics, human factors, multimedia computing, AI and NLP, and more recently education. He is a Fellow of the BCS, The Chartered Institute for IT and a Vice President of the Association of British Chinese Professors (ABCP).
\end{IEEEbiography}

\end{document}